\documentstyle[12pt]{article}

\begin{document}

\setcounter{page}{1}
\pagenumbering{arabic}
\pagestyle{plain}

\newcommand{\be}{\begin{equation}}
\newcommand{\ee}{\end{equation}}
\newcommand{\ds}{\displaystyle}
\newcommand{\bdm}{\begin{displaymath}}
\newcommand{\edm}{\end{displaymath}}

%===============================================================
\centerline {\bf  Wild cables and survivability of macroscopic}
\centerline {\bf molecular structures in hot tokamak plasmas}
\vskip 0.3cm
\centerline {\it A.B. Kukushkin, V.A. Rantsev-Kartinov }
\vskip 0.2cm
\centerline {\it INF RRC ``Kurchatov Institute'', Moscow, 123182, Russia} 
\centerline {\it kuka@qq.nfi.kiae.su, rank@qq.nfi.kiae.su} 
\vskip 0.2cm

Wild cables and survivability of macroscopic
molecular structures in hot tokamak plasmas

The evidences for tubular rigid-body structures are found in
tokamak plasmas, which are similar to long-living filaments
observed in a Z-pinch ([1] Kukushkin, Rantsev-Kartinov, Proc.
26-th EPS conf., http://epsppd.epfl.ch/cross/p2087.htm).  These
structures are suggested to be a "wild cables" produced by the
channelling of EM energy pumped from the external electric
circuit and propagated to the plasma core in the form of the
high frequency EM waves along hypothetical (carbon) microsolid
skeletons [1] which are assembled during electric breakdown. It
is shown that such skeletons may be protected from the
high-temperature ambient plasma by the TEM waves produced thanks
to the presence of microsolid skeletons.

\vskip 0.2cm

%===============================================================
{\bf 1. Introduction}.  

Recently the anomalously high survivability of some filaments in
laboratory plasmas was illustrated [1(a)] with tracing the
history of the typical long rectilinear rigid-body block in a
Z-pinch. The pictures were taken in visible light at different
time moments from different positions, during about half a
microsecond, that is comparable with the entire duration of the
Z-pinch discharge (see Fig. 1 in [1(a)]). The original images
were processed with the help of the method [2(a,b)] of
multilevel dynamical contrasting (MDC) of the images.

The phenomenon of long-living filaments (LLFs) [1,2] in various
laboratory plasmas (gaseous Z-pinches [2(a,c)), plasma foci
[2(e)], and tokamaks [2(d)]) has lead us to a conclusion [1(a)]
(see also references therein) that only the quantum (molecular)
long-range bonds inside LLFs may be responsible for their
observed survivability, rather than the mechanisms of a
classical particles plasma. Specifically, the carbon nanotubes
have been proposed to be the major microscopic building blocks
of the respective microsolid component of LLFs because such
nanotubes may be produced in various electric discharges (see,
e.g., [3]).

{\bf 2. Rigid-body structures in tokamak plasmas.} 

An analysis of available databases carried out with the help of
the MDC method [2(a,b)], shows the presence of tubular
structures. The typical examples for tokamaks TM-2, T-4, T-6 and
T-10 (major radius $R = 0.4,~0.9,~0.7,~1.5~m$, minor radius
$a = 8,~20,~20,~33~cm$, toroidal field $B_T = 2,~4.5,~0.9,
~3~T$, total current $I_p \sim 25,~200,~100, 
~300~kA$, electron temperature $T_e(0) \sim 0.6,~3,~0.4, 
~2~keV$, electron density $n_e(0) \sim (2,~3,~2,~3)
\times 10^{13}cm^{-3}$, respectively)  
are given in [4]. The figures presented there 
are taken in visible light with
the help of a strick camera and high-speed
camera. The effective time exposure is about $10~\mu sec$.  The
major features of the structuring are as follows:

(a) the length scale of the rigid-body tubular structuring
varies in a broad range, from comparable with the minor radius
of a tokamak to less than millimeter scale;

(b) the typical tubule seems to be a cage assembled from the
(much) thinner, long rectilinear rigid-body structures which
look like a solid thin-walled cylinders; 

(c) the (almost rectilinear) tubules form a network which starts
at the farthest periphery and is assembled by the tubules of
various directions;

(d) a radial sectioning of the above network is resolved which
looks like a distinct heterogeneity at a certain magnetic flux
surface(s) (such a sectioning was suggested [1(b),2(d)] to cause
the observed internal transport barriers in tokamaks).

The pictures include, in particular, the periphery of the
tokamak T-10 plasma illuminated by the carbon pellet emission
(the pellet track is outside the picture).  The system of
concentric circles and the inner almost rectilinear tubule
located approximately on the axis of these circles form together
a sort of the squirrel's wheel. Major axis of this system is
directed nearly orthogonal to toroidal magnetic field. The
system is 5 cm long and of $4\div4.5~cm$ diameter.  The central
and boundary vertical tubules are of $4~mm$ diameter. 
Similar structures appear to form in all tokamaks, i.e. with no 
regard to pellet injection.  

{\bf 3. Probable mechanism of formation and survivability of
microsolid skeletons in tokamak plasmas.}

(i) A deposit of carbon nanotubes, of the relevant quantity, is
produced at the inner surface of the chamber during discharge
training, from either graphite-containing construction elements
(like, e.g. limiters or walls) or carbon films produced by the
deposition of the organic oils normally used in the vacuum pumping
systems (the nanotubes may form due to rolling up of
monolayers ablated from solid surfaces or thin films).

(ii) Electrical breakdown occurs along chamber's surface (or its
part, namely, the inner side of the torus) and is based on the
substantially enhanced rate of (cold) autoemission and
thermoelectric emission of electrons by the nanotube (as
compared to macroscopic needles).

(iii) The microsolid skeletons are assembled from individual
nanotubes which are attracted and welded to each other by the
passing electric current to produce self-similar tubules [1(a)] of
macroscopic size, of centimeter length scale and larger (this
electric current is produced by the poloidal magnetic field
$B_{pol}$ pumped from the external circuit into the chamber). 

(iv) Once the skeleton (or its relevant portion) is assembled,
the substantial part of the incoming $B_{pol}$ brakes at it and
produces a cold heterogeneous electric current sheath made of
conventional plasma. A part of $B_{pol}$ near the skeleton is
bouncing along its every rectilinear section (i.e. between the
closest points of the deviation, even small enough, from
rectilinearity). This produces a high-frequency EM wave which,
in turn, produces, by the force of the high-frequency (HF)
pressure [5] (sometimes called in literature the Miller force),
the cylindrical cavities of a depleted electron density
(primary channels) around the skeletons. 

(v) At the skeleton's (and plasma column) edge the bouncing
boundary of the cavity from the scrape-off layer side produces a
HF valve for the incoming $B_{pol}$, because of the node of the
standing wave at the edge.  This works as a HF convertor of a
part of the incoming $B_{pol}$ which is transported then along
the skeleton in the form of EM waves. (Besides, a part of
$B_{pol}$ which reaches the cavity in the conventional regime of
the diffusion of $B_{pol}$, is transformed into a HF field by
the oscillating boundary of the cavity).  The EM waves sustain
the cavity and protect the skeletons from direct access of
thermal plasma particles.  Therefore the skeleton appears to be
an inner wire of the cable network (a {\it wild cable} network)
in which the role of a screening conductor is played by the
ambient plasma.

In this paper, we restrict ourselves to quantifying the above
picture in its quasi-stationary stage of energy inflow
through the wild cable network. 

For the frequency $\omega_c$ of the major harmonic of EM
oscillations trapped in radial direction in a 
cylindrical almost-vacuum cavity of
effective radius $r_c$ around microsolid tubule of length $L_c$,
one has ($\omega_{pe}$ is plasma frequency, $c$, the speed of 
light):
\be			\label{lambda}
\omega_c \simeq \pi c/L_c \leq \omega_{pe}.  
\ee
\noindent 
For tokamak geometry, one has the following chain of
transformations of EM waves. The cavities at plasma edge (they
normally possess some declination with respect to the boundary
magnetic surface) allows the field lines of $B_{pol}$ to move
directly inside the cavity and thus produce the magnetic (H)
wave. For the strongest EM wave among H waves, $H_{11}$ wave, 
one has: 
$\lambda \simeq 2L_c \geq \lambda_{crit} \sim \alpha r_c$, where
$\lambda_{crit}$ is the critical wavelength for free propagation of 
the respective EM wave in the cable ($\alpha_{H11} \approx \pi$).  
Therefore, the trapping of $H_{11}$ wave
in the edge cavity leads to the wiring of magnetic field lines
round the inner wire that produces TEM and electric (E) waves
propagating in both directions (the strongest wave among E-waves
will be $E_{01}$ wave). However, the $E_{01}$ wave will also be
trapped in the cavity ($\alpha_{E01} \approx 2.6 $), in
contrast to TEM wave ($\lambda_{crit}^{TEM} = \infty$).  Also,
the H and E-waves, in contrast to TEM wave, are detached from
the wall (in radial direction, these waves are the standing
ones) so that only the TEM wave can actually maintain the
boundary of the cavity.  Thus, the edge cable converts a part of
$B_{pol}$ into HF TEM wave propagating inward. The signs of this
HF field of which a small part is reflected outward may be found
in the measurements of EM fields outside plasma column (see
below).

It is assumed also that the presence of an external stationary
strong magnetic field doesn't influence substantially the form
of the cavity, because even for $\omega_c \ll
\omega_{He}$ ($\omega_{He}$ is electron gyrofrequency) 
the amplitude $\vec E_0$ of the HF electric field may have a
non-zero component parallel to magnetic field (in that case we
will assign $\vec E_0$ to the respective component of the
amplitude).

The distribution of plasma density around the inner wire can be
described by a set of equations for the two-temperature
quasi-hydrodynamics of a plasma in a HF EM field [6].
Under condition $l_E \gg r_D$, where $l_E$
is the characteristic length of spatial profile of 
$E_0(\vec r)$ and $r_D$ is Debye radius, 
one can neglect the deviation from quasi-neutrality  and arrive at
quasi-Boltzmann distribution (see e.g. [6(b)]): 
$n_e =  n_{e0}~ \exp(-\Psi/(T_e+T_i) )$, where 
$\Psi = e^2E_0^2/( 4m_e\omega_c^2),~$ 
$n_{e0}$ is background density of plasma electrons. 
The condition for plasma detachment from the inner wire reads: 
\be \label{Emin} 
eU_0 \geq 2\pi (r_c/L_c) \sqrt{A m_e c^2 (T_e+T_i)}, ~~~
A \sim (r_w^2/r_c^2)~ln(n_{e0}/n_{emin}),    
\ee
\noindent
where $U_0$ is the effective voltage bias of the TEM wave in the
cable ($E_0(r) \sim U_0/r$; $r$ is the radial coordinate in a
circular cylindrical cable, $r_w$, radius of inner wire), 
$n_{emin}$ is the minimal density
permitted, at a temperature $T_e$, for the inner wire to be not
destroyed by the plasma impact. For tokamak case ($n_{e0} \sim
10^{13}cm^{-3}$), we take $A \sim 5$.  

Equation (\ref{Emax}) is to be coupled to the condition of
the applicability of the concept of the $(- \nabla \Psi)$ force,
$\rho \ll l_E$ ($\rho$ is the amplitude of electron's
oscillations in the HF electric field).
For our
estimates, this limitation, however, may be weakened and
takes the form:
\be \label{Emax} 
eU_0 \leq \pi^2~m_ec^2~r_c(r_c-r_w)/L_c^2,
\ee
And finally, the HF electric field in the cables may be related
to the observable turbulent electric fields because wild
cables are the strong sources of electrostatic oscillations in
plasma. As far as there should be a sort of
the feedback between plasma and cavity, one may consider the
cable's cavity as a soliton with such a strong reduction of the
eigenfrequency (a redshift) that soliton's velocity becomes
independent from dispersion. For $W/nT \leq 1$ ($W \equiv
E_0^2/16\pi$) this gives rough estimate:
\be \label{W/nT} 
W/nT \sim (1- (\omega_c/\omega_{pe}) ).    
\ee
\noindent
At the quasi-stationary stage of discharge, one may evaluate the
spatial distribution of the amplitude $E_{turb}$ of the
turbulent electric field, regardless of its spectral
distribution, as being described, in radial direction with respect to
the individual cable, by the scaling law of the TEM wave. For the
contribution of a single cable, one has:
\be \label{Eturb} 
E_{turb}(r) \sim U_0/r. 
\ee
\noindent

Equations (\ref{lambda}), (\ref{Emin}) and (\ref{Emax}), along
with rough estimates of Eqs.(\ref{W/nT}) and (\ref{Eturb}),
establish a set of equations that enable one to evaluate the
plausibility of the presence of wild cables in tokamak plasmas,
using available data on measuring the values of $\omega_c$ [7]
(and/or $L_c$) and $E_{turb}$ [8].

Now we can test the problem for typical data from the periphery
of the T-10 tokamak, keeping in mind the closeness of T-10 regimes 
analyzed in [7,8] and those for Figure 4.  
First, the spectra of the HF EM field in
the gap between the plasma column and the chamber measured in
the GHz frequency range revealed [7] a distinct bump at $\nu_c
\sim (4\div5) \times 10^9 Hz$, of the width $\sim 2 \times 10^9 Hz$, 
which always exists in ohmic heating regimes and increases with
electron cyclotron heating (this bump is a stable formation and
it moves to the lower frequencies and turns into a peak only
under condition of strong instabilities, especially disruption
instability). This gives $L_c \approx 3~cm$. Note that this is
in reasonable agreement with the data from  the high-speed camera
picture for T-10 plasma periphery where $L_c \approx 4\div5~cm$.

Second, the analysis of observations of Stark broadening of
deuterium spectral lines (and their polarization state) at the
periphery of the T-10 tokamak in the region of $T_e \sim 100~
eV$, allowed [8] to estimate the spectral range of HF electric
fields ($\omega \approx \omega_{pe} \sim 10^{11}Hz$), their
amplitude ($E \sim 10\div20~kV/cm$) and angular distribution.

For $L_c = 3~cm, T_e = T_i = 100~eV$, Eqs. (\ref{Emin}) and
(\ref{Emax}) give a constraint $ S \equiv (r_c-r_w)/L_c \geq
0.03~ $. For $(r_c-r_w)
\sim r_c$,  from Eq. (\ref{Emin}), 
one can find the absolute minimum of voltage bias:  
$(U_0)_{min} \approx 5~kV$. For $S = 0.03$, Eqs. (\ref{Emin}) 
and (\ref{Emax}) give $U_0 \approx 5~kV$, while for $S = 0.1$
one has  $ 15 \leq  U_0~(kV) \leq 50$.  
Further, Eq. (\ref{W/nT}) gives $E_0(r_c) \geq 50~kV/cm$, while, 
for $r_c \sim 1\div2~mm$ and $<r> \sim 1\div3~cm$ ($<r>$ is 
the average distance between individual cables in the region of
observation), Eq. (\ref{Eturb}) gives the estimate $E_0(r_c)
\geq 10^2~kV/cm$, or $U_0 \geq 10~kV$.   
The results of numerical solution of the Poisson equation [6]
show that, e.g., for $U_0 = 30~kV$ at the distances $r \sim
2\div3~mm$ the plasma density falls down, with respect to its
background value, by the seven-eight orders of magnitude.

{\bf 4. Conclusions}

The experimental data of Sec. 2 and the model of Sec. 3 support
the hypothesis [1] that plasmas with long-living filaments is
such a form of the fourth state of matter, which is an intricate
mixture of three other states (gaseous, liquid and solid).  The
presence of the inner wire (namely, electrically conducting
microsolid skeleton) in the wild cable is responsible not only
for the observed anomalous mechanical stability of this
structure but also for the formation of TEM waves in the cavity
that is critical for the self-sustainment of the cavity and for
the transport of EM energy to plasma core.

It follows that observed structuring could be: 

(i) a strong candidate for the nonlocal (non-difusion) component
of heat transport (and observed phenomena of fast nonlocal
responses) in tokamaks; 

(ii) a powerful source of non-linear waves and (strong)
turbulence throughout plasma volume;

(iii) a low-dissipation waveguide responsible for the spatial
profile of poloidal magnetic field in tokamaks, rather than
total resistance of plasma (in agreement with the observed
applicability of Spitzer, or close, resistivity to describing
the ohmic heat release in plasma);

(iv) a universal phenomenon in well-done laboratory plasmas and
space; in particular, similar wild cables may form in gaseous
and wire-array Z-pinches and be responsible for the fast
nonlocal transport of EM energy toward Z-pinch axis.

{\bf Acknowledgments.} 

The authors are indebted to V.M.  Leonov, S.V.  Mirnov and I.B.
Semenov, K.A. Razumova, and V.Yu.  Sergeev for presenting the
originals of the data from tokamaks T-6, T-4, TM-2, and T-10,
respectively.  The authors appreciate discussions of the paper
with V.V. Alikaev, V.I. Poznyak and V.L. Vdovin, and
participants of seminars in the Institute of Nuclear Fusion.
Our special thanks to V.I. Kogan for his interest and support,
and V.D. Shafranov, for valuable discussion of the paper.

\vskip 0.2cm

\centerline{REFERENCES}

[1]
%\bibitem{EPS'99} 
Kukushkin A.B., Rantsev-Kartinov V.A.,
Proc. 26-th Eur. Phys. Soc. conf. on Plasma Phys. and Contr.
Fusion, Maastricht, Netherlands, June 1999, (a) p. 873
(http://epsppd.epfl.ch/cross/p2087.htm); (b) p. 1737
(p4096.htm).

[2]
%\bibitem{all-LLF} 
Kukushkin A.B., Rantsev-Kartinov V.A., (a) Laser and Part. Beams,
{\bf 16}, 445 (1998); 
(b) Rev. Sci. Instrum.,  {\bf 70}, 1387 (1999); (c) Ibid, p. 1421; 
(d) Ibid., p. 1392; (e) 
Kukushkin A.B., et. al., Fusion Technology, {\bf 32}, 83 (1997). 
[3]
%\bibitem{Eletskii}
Eletskii A.V., Physics-Uspekhi, {\bf 167}, 945 (1997). 
[4] 
Kukushkin A.B., Rantsev-Kartinov V.A., Preprint of the RRC
Kurchatov Institute, IAE-6157/6, Moscow, October 1999
(submitted to JETP Lett.). 

[5]
%\bibitem{HF-force} 
Gaponov A.V., Miller M.A., Zh. Exp. Teor. Fiz. (Sov. Phys.
JETP), {\bf 34}, 242 (1958); Volkov T.F., In: Plasma Physics and
the Problem of Controlled Thermonuclear Reaction, Ed.
M.A.Leontovich, [In Rus.], USSR Acad.Sci., 1958, Vol. 3, p.
336, Vol. 4, p. 98; Sagdeev R.Z., Ibid., Vol.3, p. 346.  
  
[6] 
%\bibitem{HF-hydro}
(a) Gorbunov L.M., Uspekhi Fiz. Nauk (Sov. Phys. Uspekhi), {\bf 109}, 
631 (1973);
(b) Litvak A.G., In: Voprosy Teorii Plazmy (Reviews of Plasma Phys.), 
Eds. M.A.Leontovich and B.B.Kadomtsev, [In Rus.], vol. 10, p. 164.

[7]
%\bibitem{HF-spectrum}
Poznyak V.I., et. al. Proc. 1998 ICPP and 25-th Eur. Phys. Soc.
Conf. on Plasma Phys. and Contr.  Fusion, 1998, Prague, ECA Vol.
22C (1998) p. 607.

[8]
%\bibitem{HF-ampl}
Rantsev-Kartinov V.A., Fizika Plazmy (Sov. J. Plasma Phys), {\bf
14}, 387 (1987); Gavrilenko V.P., Oks E.A., Rantsev-Kartinov
V.A., Pis'ma Zh. Exp. Teor. Fiz. (JETP Lett.), {\bf 44}, 315 
(1987).

\end{document}